\documentclass[
reprint,
superscriptaddress,
 amsmath,amssymb,
aip,
pop,
floatfix,
]{revtex4-1}

\usepackage{graphicx}
\usepackage{epstopdf}
\usepackage{dcolumn}
\usepackage{bm}
\usepackage{ifpdf}
\usepackage[usenames,dvipsnames]{color}
\usepackage{hyperref}
\def\Snospace~{\S{}}

\usepackage[all]{hypcap}
\hypersetup{urlcolor=Blue, citecolor=Blue,linkcolor=Blue,colorlinks=true}

\newcommand{\Fig}[1]{
		\label{fig:#1}
		\ifpdf
			\includegraphics[scale=1]{pop_2017_figures_#1-eps-converted-to.pdf} 
		\else
			\includegraphics[scale=1]{pop_2017_figures_#1.eps}
		\fi
}

\newcommand{\x}[1]{$x=#1$ mm}
\newcommand{\y}[1]{$y=#1$ mm}
\newcommand{\z}[1]{$z=#1$ mm}
\renewcommand{\t}[1]{$t=#1$ ns}

\newcommand{\xcmcubed}[2]{$#1\times10^{#2}$ cm$^{-3}$}
\renewcommand{\deg}[1]{#1$^{\circ}$}
\newcommand{\zte}{\bar{Z}T_e}
\newcommand{\sref}[2]{\hyperref[#1]{Fig. \ref{#1}#2}}
\DeclareMathOperator{\sech}{sech}

\begin{document}

\title{Formation and Structure of a Current Sheet in Pulsed-Power Driven Magnetic Reconnection Experiments}

\author{J. D. Hare}
\email{jdhare@imperial.ac.uk}
\author{S. V. Lebedev}
\email{s.lebedev@imperial.ac.uk}
\author{L. G. Suttle}
\affiliation{Blackett Laboratory, Imperial College, London, SW7 2AZ, United Kingdom}
\author{N. F. Loureiro}
\affiliation{Plasma Science and Fusion Center, Massachusetts Institute of Technology, Cambridge MA 02139, USA}
\author{A. Ciardi}
\affiliation{Sorbonne Universit\'{e}s, UPMC Univ Paris 06, Observatoire de Paris, PSL Research University, CNRS, UMR 8112, LERMA, F-75005, Paris, France}
\author{G. C. Burdiak}
\altaffiliation[Current address: ]{First Light Fusion Ltd, 10 Oxford Industrial Park, Yarnton, Kidlington OX5 1QU, UK}
\author{J. P. Chittenden}
\author{T. Clayson}
\author{S. J. Eardley}
\author{C. Garcia}
\author{J. W. D. Halliday}
\author{N. Niasse}
\altaffiliation[Current address: ]{First Light Fusion Ltd, 10 Oxford Industrial Park, Yarnton, Kidlington OX5 1QU, UK}
\author{T. Robinson}
\author{R. A. Smith}
\author{N. Stuart}
\author{F. Suzuki-Vidal}
\author{G. F. Swadling}
\altaffiliation[Current address: ]{Lawrence Livermore National Laboratory, California 94550, USA}
\affiliation{Blackett Laboratory, Imperial College, London, SW7 2AZ, United Kingdom}
\author{J. Ma}
\affiliation{Northwest Institute of Nuclear Technology, Xi'an 710024, China}
\author{J. Wu}
\affiliation{Xi'an Jiaotong University, Shaanxi 710049, China}

\date{\today}

\begin{abstract}
	We describe magnetic reconnection experiments using a new, pulsed-power driven experimental platform in which the inflows are super-sonic but sub-Alfv\'enic.
	The intrinsically magnetised plasma flows are long lasting, producing a well-defined reconnection layer that persists over many hydrodynamic time scales.
	The layer is diagnosed using a suite of high resolution laser based diagnostics which provide measurements of the electron density, reconnecting magnetic field, inflow and outflow velocities and the electron and ion temperatures.
	Using these measurements we observe a balance between the power flow into and out of the layer, and we find that the heating rates for the electrons and ions are significantly in excess of the classical predictions.
	The formation of plasmoids is observed in laser interferometry and optical self-emission, and the magnetic O-point structure of these plasmoids is confirmed using magnetic probes.
\end{abstract}

\maketitle
\section{Introduction}

Magnetic reconnection plays a fundamental role in plasma physics, as it provides a mechanism to break the frozen-in condition which constrains the magnetic field and the plasma to move together\cite{Yamada2010,Zweibel2016}.
As such, reconnection is a powerful influence on the dynamics and structure of the plasma which makes up the majority of the visible matter in the universe.

Active areas for investigation include the rate of reconnection, the partitioning of the magnetic energy between the electrons and ions, and the formation of instabilities during the reconnection process.
Reconnection has been studied in a range of physical systems, including observations of solar flares, in-situ measurements of the magnetosphere by satellites, in fusion-capable laboratory devices and in dedicated laboratory reconnection experiments (see Refs. \onlinecite{Yamada2010, Zweibel2016} and references therein).

Dedicated laboratory studies include the use of magnetically confined plasmas and colliding, laser produced plasmas.
In magnetically driven experiments, such as MRX,\cite{Ji1999,Yamada2015a,Yamada2016} TS-3,\cite{Ono2011} and TREX,\cite{Olson2016}, a long lasting reconnection layer forms from inflows in which the magnetic energy is much larger than the thermal or kinetic energies, resulting in inflows which are sub-sonic and sub-Alfv\'enic.
These experiments are usually diagnosed with in-situ probes which measure the magnetic field and the electron density and temperature, and spectroscopic measurements are used to measure the ion temperature.

In laser-driven experiments, the plasma is produced by laser-solid interactions\cite{Nilson2006,Fiksel2014b}, and thin ribbons of magnetic flux are transiently annihilated between two expanding bubbles.
In these experiments, the inflows are significantly super-sonic and super-Alfv\'enic, with the thermal and kinetic energies typically far larger than the magnetic energy --- as such, heating due to magnetic reconnection is often much smaller than the thermalisation of the flow kinetic energy.\cite{Fox2011,Rosenberg2012}
Laser driven experiments have been diagnosed using Thomson scattering and proton radiography.

In this paper we present a recently developed pulsed-power driven experimental platform,\cite{Suttle2016,Hare2017} which allows reconnection to be observed in a regime in which the magnetic, thermal and kinetic energies are in rough equipartition in the inflowing plasma.
This platform converts initially solid wires into plasma using an intense pulse of electrical current, and then accelerates this plasma to high velocity.

The regime of reconnection depends on the choice of wire material. 
In previous work, we have described super-sonic, super-Alfv\'enic inflows formed from aluminium wires \cite{Suttle2016}, in which we saw the formation of standing shocks and the pile up of magnetic flux outside the flux annihilation region.
In the aluminium experiments, the magnetic energy was a significant (but not the dominant) source of energy in the inflows, radiative cooling was an important loss mechanism, and plasmoids were not observed inside the smooth interaction layer, despite the presence of the same density perturbations as in the carbon experiments.
In this paper we describe experiments with carbon wires, which produce super-sonic but sub-Alfv\'enic inflows, which did not produce shocks or lead to the pile up of magnetic flux.
The magnetic energy formed the dominant contribution to the overall power balance, with negligible radiative cooling, and the layer was highly unstable to the formation of plasmoids.\cite{Hare2017, Hare2017a}

These experiments focus on symmetric reconnection, with no guide field and open boundaries for the outflows.
We describe the formation of a quasi-2D, elongated ($L/\delta\sim10$, where $L$ and $\delta$ are the half-length and half-width of the layer, respectively) and long lasting reconnection layer in approximate pressure balance. 
We observe high electron and ion temperatures, a Harris sheet-type magnetic field profile and the formation of plasmoids within the reconnection layer. 
Using these results, we show the heating rates within the reconnection layer are significantly faster than predicted for classical mechanisms, and we discuss the electric field balance across the reconnection layer, which cannot be explained by steady-state resistive MHD

\begin{figure*}[t]
	\Fig{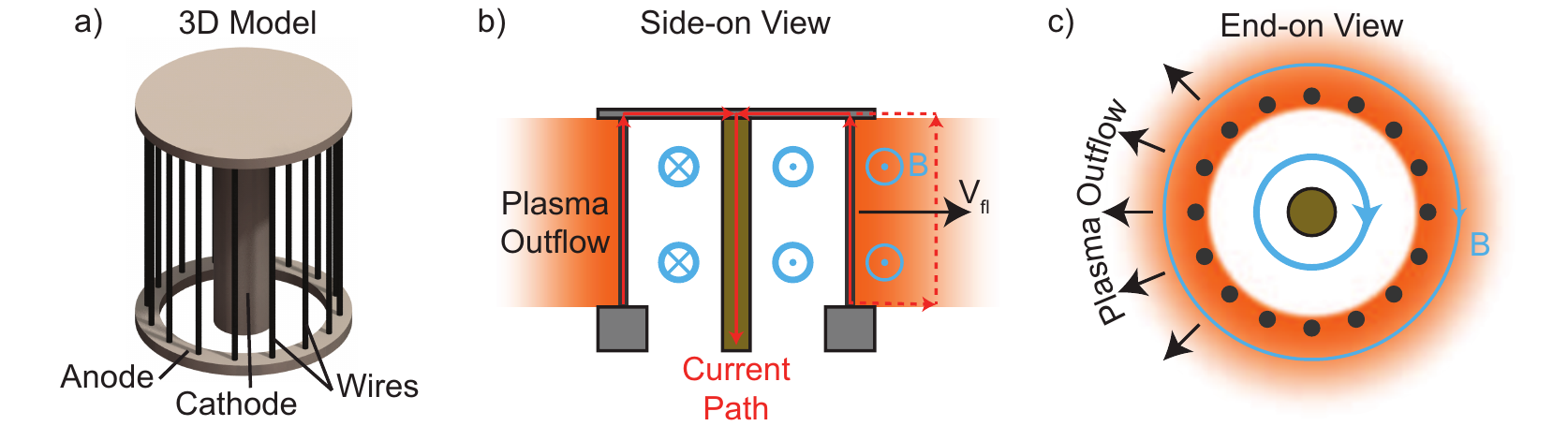}
	\centering
	\caption{Schematic for an exploding (`inverse') wire array. a) 3D diagram, showing the cylindrical arrangement of thin wires surrounding a central cathode. b) Cut-away view from the side, showing the current path at the beginning of the current pulse (in red), the azimuthal magnetic field (blue), the acceleration of the outflows to the flow velocity $V_{fl}$ and the advection of the magnetic field, with the new current path shown (dashed red line). c) Cut-away view from the top, showing the large azimuthal magnetic field around the central conductor and the convection of a fraction of this magnetic field by the plasma flows.}
\end{figure*}

This paper is organised as follows: In \autoref{sec:experimental_setup} we discuss the driver and plasma source for these experiments, as well as the reconnection geometry used in this paper. 
The diagnostic suite for these experiments is discussed in \autoref{sec:diagnostics}, which includes laser interferometry, Thomson scattering, Faraday rotation imaging and magnetic probes. 
The results from these diagnostics are presented in \autoref{sec:results}, and the results are discussed in detail in \autoref{sec:discussion}. 
We conclude and present the outlook for future work in \autoref{sec:conclusions}.

\section{Experimental Set-up}\label{sec:experimental_setup}

Experiments were carried out on the MAGPIE generator at Imperial College London.\cite{Mitchell1996}
The current pulse is well approximated by $I=I_0\sin^2\left(\pi t/2\tau\right)$, where the peak current $I_0=1.4$ MA and the rise time $\tau=240$ ns. This current pulse delivers a peak power of 1 TW into a volume of a few cubic centimetres.

The experiments described in this paper use exploding (or `inverse') wire arrays\cite{Harvey-Thompson2009} as plasma sources.
An exploding wire array consists of a cylindrical cage of thin wires which surround a thick central conductor (\hyperref[fig:exploding_wire_array]{Fig. 1a}).
The base of the cage is attached to the anode of the \textsc{MAGPIE} generator, and the central conductor is attached to the cathode.
The generator drives an electrical current pulse up through the wires, along the current path shown in \hyperref[fig:exploding_wire_array]{Fig. 1b}, which is a cut-away view from the side.
The current heats and ionises the wire material, producing a coronal plasma which surrounds a cold, dense wire core.\cite{Lebedev2001}
A fraction of the drive current flows through this coronal plasma, and this interacts with the large azimuthal magnetic field surrounding the central conductor, producing a $\mathbf{J}\times\mathbf{B}$ force radially outwards.
Initially the individual wires are surrounded by closed magnetic field lines, but as the plasma accelerates outwards it carries with it some fraction of the drive current.
This current causes a reconfiguration of the overall magnetic topology\cite{Chittenden2004, Greenly2009} that results in a global azimuthal magnetic structure.
The coronal plasma is continuously replenished by  ablation from the surface of cold wire cores and this plasma is continuously accelerated outwards for the duration of the current pulse, resulting in long lasting, magnetised plasma flows.\cite{Lebedev2014,Swadling2016}

\begin{figure}[t]
	\Fig{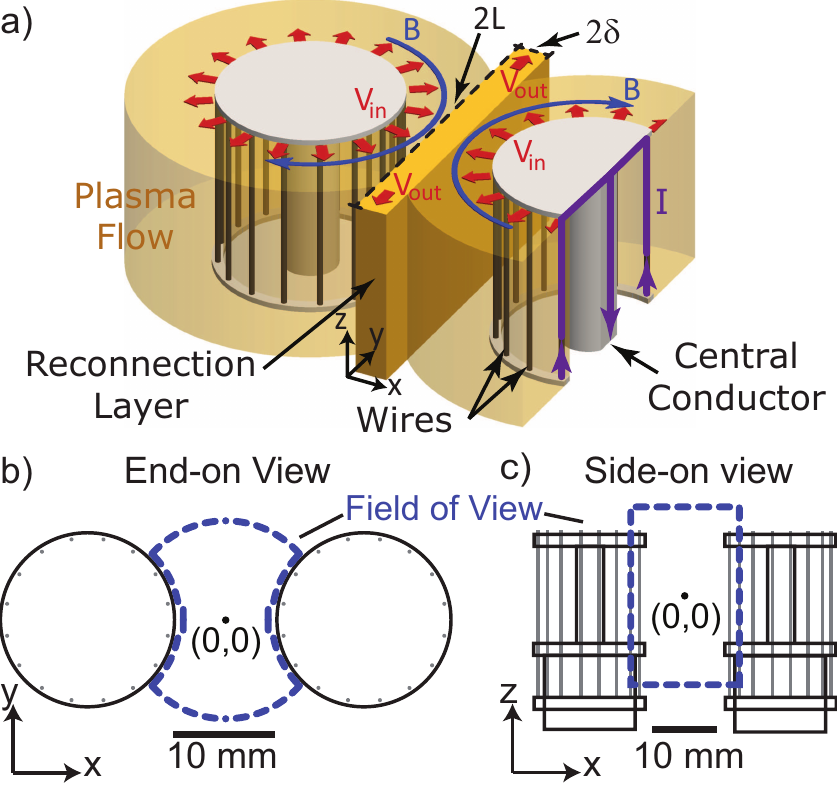}
	\centering
	\caption{a) 3D schematic of the magnetic reconnection experiment, using two exploding wire arrays driven in parallel. The right-hand array is cut-away to show the current path. The radially diverging flows from each array collide at the mid-plane where  the advected magnetic fields are anti-parallel, producing a well-defined, quasi-2D reconnection layer of width $2\delta$ and length $2L$. b) Top view, showing the field of view for the laser interferometry and optical fast-framing diagnostics. c) Side view, showing the field of view for the Faraday rotation imaging diagnostic. The co-ordinate system and origin [marked with a dot and (0,0)] is shown for each view.}
\end{figure}

The pulsed-power-driven reconnection platform\cite{Suttle2016, Hare2017} uses two exploding wire arrays (\sref{fig:3D_double_exploders}{a}), In the experiments described in this paper, each array consisted of 16 carbon wires (Staedler Mars Micro Carbon B) with a diameter of 400 $\mathrm{\mu}$m, evenly spaced around a circle with a 16 mm diameter.
The wire arrays were 16 mm tall, and were placed 27 mm apart (centre to centre).
This separation meant that the radius of curvature of the magnetic field lines at the mid-plane where the reconnection layer formed was $R_c=13.5$ mm, and we use half this value as the length scale of the reconnection layer, $L=7$ mm.

The current pulse is split evenly between the two arrays, which generate two diverging outflows, which collide at the mid-plane (\x{0}).
When the flows collide, the advected magnetic fields in the two oppositely directed flows are anti-parallel, and magnetic reconnection occurs.
This set-up is inherently symmetric, with no guide field ($B_z=0$), and open boundary conditions as the outflows expand into a large vacuum chamber.
Plasma and magnetic field are continuously injected into the reconnection layer for the duration of the current pulse, around 500 ns.

\section{Diagnostics}\label{sec:diagnostics}

The relevant plasma parameters in these reconnection experiments were measured using a suite of high-resolution, non-perturbative, temporally and spatially resolved diagnostics.
These diagnostics include laser interferometry, Thomson scattering\cite{Harvey-Thompson2009,Swadling2014a}, Faraday rotation imaging and an optical fast frame camera, which enabled measurements of the electron density, flow velocity, electron and ion temperatures and the reconnecting magnetic field.
In-situ measurements of the reconnected magnetic field and plasmoid field structure were carried out using magnetic probes in some experiments.

Two-colour, two-time laser interferometry was performed using the 2nd and 3rd harmonics of a Nd-YAG laser (EKSPLA SL321P, 500 ps, 500 mJ) at 532 nm and 355 nm.\cite{Swadling2014a}
The 355 nm pulse propagates along the same optical path as the 532 nm pulse, delayed by 20 ns to provide two frames of interferometry. 
The beams were imaged onto Canon 350D and 500D DSLRs, with the time resolution set by the length of the laser pulse.
The interferometry used a Mach-Zehnder set-up to provide measurements of the line-integrated electron density ($\int n_e dz$), and in these experiments the probing was performed from the top, looking down onto the reconnection plane (\hyperref[fig:3D_double_exploders]{Fig. 2b}).
The interferograms were processed using a custom software suite\cite{Swadling2013} which removes the flat features caused by Delaunay triangulation of the isophase contours.\cite{Swadling2014a}

Thomson scattering was carried out using a focused laser beam (8 ns, 3 J, 532 nm) which passed through the reconnection layer.
The scattered light was imaged using two sets of fibre optic bundles placed on opposite sides of the vacuum chamber.\cite{Harvey-Thompson2009,Swadling2014a}
Each fibre optic bundle consisted of 14 fibres, and the position of the bundle and the imaging lens set the spatial range over which scattered light was collected.
The bundle positions were chosen such that the two bundles were sensitive to orthogonal velocity components\ (see Ref. \onlinecite{Swadling2014} and \hyperref[fig:thomson_scattering]{Fig. 6}) , which were aligned with the inflow ($x$) and outflow ($y$) directions.
The spectra were dominated by the ion feature of the scattered light, and fit using a custom code\cite{Froula2011, Hare2017a} which determined the flow velocity ($V_x$ or $V_y$), the ion temperature ($T_i$)  and the product of the effective charge and the electron temperature $\bar{Z}T_e$.
A non-Local-Thermodynamic-Equilibrium (nLTE) code\cite{Chittenden2016} was used to decompose $\bar{Z}T_e$ into $\bar{Z}$ and $T_e$.

Faraday rotation imaging was performed using an infra-red laser beam (1053 nm, 5 J, 1 ns)\cite{Swadling2014a} which passed side-on (\hyperref[fig:3D_double_exploders]{Fig. 2c}), along the direction of the reconnecting magnetic field.
The polarisation of the probing laser beam rotates due to the different phase velocities of the left and right circularly polarised waves, giving the angle of rotation:
\begin{equation}
\alpha=\frac{e^3\lambda^2}{8\pi\epsilon_0m_e^2c^3}\int n_e B_y dy.
\end{equation}
The laser beam is split to provide an in-line Mach-Zehnder interferometer, and then passes into the polarimetry diagnostic, which splits the laser again and images it onto two identical CCDs (Atik 383L+) through polarisers set to $\pm$\deg{3} from the extinction angle.
Hence when the polarisation of the laser beam rotates in one direction, that region of the laser beam becomes darker on one CCD and lighter on the other.
Using two channels allows the optical self emission of the plasma to removed, reducing systematic errors in determining the polarisation of the laser beam.
The line-averaged reconnecting magnetic field can be determined by combining the polarisation map ($\alpha$) and the line integrated electron density from the in-line interferometry diagnostic:
\begin{equation}
B_y(x,z)=\frac{8\pi\epsilon_0m_e^2c^3}{e^3\lambda^2}\frac{\alpha(x,z)}{\int n_e dy}.
\end{equation}
A detailed description of the Faraday rotation imaging diagnostic is provided in Ref. \onlinecite{Swadling2014a}.

A fast framing camera imaged optical self emission from the plasma along the same line of sight, and with the same field of view (\hyperref[fig:3D_double_exploders]{Fig. 2b}), as the end-on laser interferometry, with a low-pass filter ($>600$ nm) to remove the laser light.
The exposure time was 5 ns, with 15--20 ns between frames, and a total of 12 frames per shot.
This camera allowed the evolution of the layer to be studied over long time-scales ($>200$ ns) in a single shot.
In some experiments, magnetic probes were used in conjunction with an optical fast-framing camera in order to measure the reconnected ($B_x$) magnetic field in the plasma outflows, with probes positioned at \x{0}, \y{\pm8}.
These probes were also used to examine the magnetic field structure of the plasmoids observed by the fast framing camera.

The magnetic probes consisted of two oppositely wound loops (0.5 mm diameter) in a thin, enamel-coated wire, which was twisted along its length and protruded from thin aluminium tube, which acted as an electrostatic shield.
The potential difference across the two loops is:
\begin{equation}V_\pm=CV_P\pm A\frac{\partial B}{\partial t}\end{equation}
where $V_P$ is the plasma potential and $C$ is some coupling constant between the probe and the plasma, and $A$ is the loop area.
The potential difference was measured separately for both loops, which showed that the electrostatic component was much smaller than the inductive component, $|CV_P|\ll |A\,\partial B/\partial t|$.
The inductive component was recovered from the difference of the two probe voltages, giving:
\begin{equation}B(t)=\frac{1}{2A}\int_0^t \left[V_+(t)-V_-(t)\right]\,dt\end{equation}
The probes were placed in the field of view of the optical fast-framing camera, which allowed the interaction of the plasmoids with the probes to be correlated with the signal on the probes. 

\section{Results}\label{sec:results}

The diagnostic suite described above allows for detailed measurements of the relevant plasma parameters within this reconnection experiment.
These experiments were highly reproducible, with similar dynamics for the inflows and layer formation in every shot.
This uniformity of the reconnection layer in the out-of-plane ($z$) direction justifies using a quasi-two-dimensional treatment when analysing these results.

Typical plasma parameters obtained in these experiments are summarised in \hyperref[tab:params]{Table I}\cite{Hare2017}. The ion skin depth $d_i=c/\omega_{pi}$ and $\lambda_{ii}$ is the ion-ion mean free path (page 31 of Ref. \onlinecite{Huba2016}):
\begin{table}[h]
	\label{tab:params}
	\centering
	\caption{Plasma parameters in the inflowing plasma and reconnection layer.}
	\begin{tabular}{|c|c|r|r|r|r|r|r|r|}
		\hline
		\rule{0pt}{2.5ex}
		Parameter & \(n_e\)            & \(\bar{Z}\)& \(V_x \left(V_y\right)\) & \(B_y\) & \(T_i\) & \(T_e\)     & \(c/\omega_{pi}\) & \(\lambda_{ii}\)\\ 
		\textit{Units}     & (cm\(^{-3}\))       & & (km/s)                  & (T) & (eV)  & (eV)      & (\(\mu\)m)          & (\(\mu\)m) \\ 
		\hline
		Inflow    & \(3\times10^{17}\) &4& 50\hphantom{)}                      & 3     & 50 & 15 & 700               & 3                 \rule{0pt}{3ex}  \\ \hline
		Layer     & \(6\times10^{17}\) &6& (130)                &    0   & 600     &100         & 400               & 30 \rule{0pt}{3ex}  \\ 
		\hline
	\end{tabular}
\end{table}

Next we presents details of the measurements of the various plasma parameters in the different stages of the reconnection process.

\subsection{Electron density maps from laser interferometry}

\begin{figure*}[!t]
	\Fig{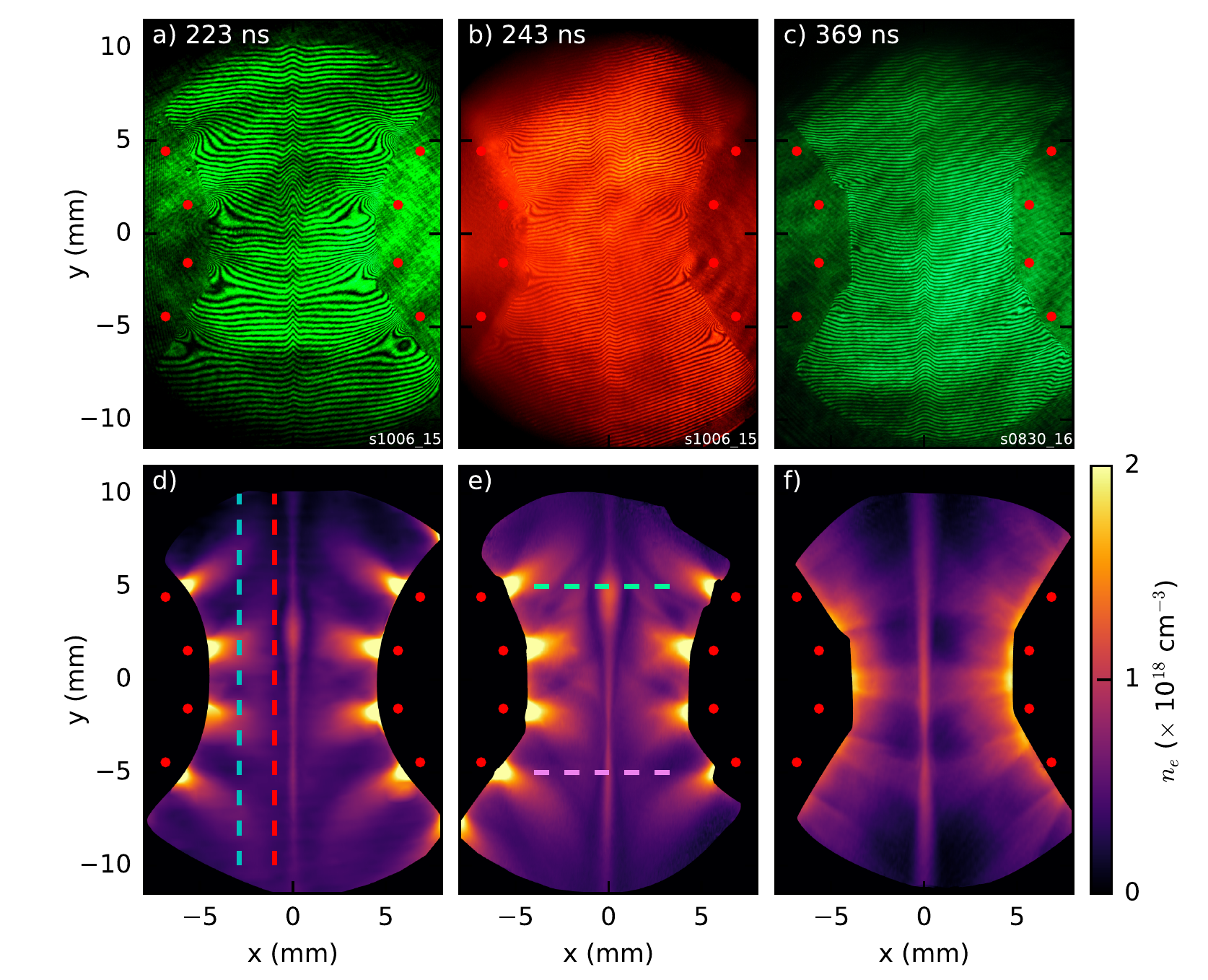}
	\centering
	\caption{Laser interferograms and processed electron density maps showing the evolution of the reconnection layer. a)-c) Raw interferograms taken at a) \t{223}, b) \t{243} and c) \t{369} after current start. The location of the wire cores is marked with red circles. d)-f) Processed electron density maps corresponding to the interferograms in a)-c). Dashed lines in d) and e) mark the position of line-outs shown in \hyperref[fig:electron_density_lineouts]{Fig. 4a and b} respectively.}
\end{figure*}

Laser interferometry shows that the reconnection layer forms at around \t{190} after current start, and the layer exists throughout the current pulse until at least \t{370}.
A typical end-on interferogram (looking down onto the reconnection plane, with the field of view indicated in \hyperref[fig:3D_double_exploders]{Fig. 2b}) from \t{223} after current start is shown in \hyperref[fig:electron_density_maps]{Fig. 3a}, in which the layer is fully formed.
A second interferogram taken 20 ns during the same shot, at \t{243}, is shown in \sref{fig:electron_density_maps}{b}, and shows the evolution of the structure of the reconnection layer.
The final interferogram in \sref{fig:electron_density_maps}{c} was taken in a different shot at \t{369}, and shows that the layer persists for almost 200 ns.

These interferograms have been processed to produce three electron density maps (\hyperref[fig:electron_density_maps]{Figs. 3d, 3e and 3f}).
The electron density is obtained by dividing the areal electron density by the length scale of the plasma that the probing beam passes through, $h\approx 16$ mm.
The location of the wire cores are marked with red dots, and the ablation flows from each wire are clearly visible from the two arrays on the left and right hand side.
The reconnection layer has already formed at the mid-plane at \t{223}, and extends over the entire field of view (21 mm) with a width of $\delta\approx0.6$ mm.
Although the density does increase inside the layer, there is no evidence for a sharp density jump caused by the formation of shocks, suggesting that the inflows in this experiment are sub-Alfv\'enic, which was confirmed using measurements of the magnetic field and plasma flow velocities discussed below.

The layer is narrow, elongated and uniform over most of its length, with the exception of a prominent elliptical region of enhanced density, which we call a plasmoid.
The plasmoid is wider and denser than the rest of the layer, and has formed inside the layer.
We will discuss these plasmoids in more detail in \autoref{ssec:plasmoids}.
The plasmoid is visible in both electron density maps ---  at \t{223} it is at \y{2.5}, and it has moved to \y{5} by \t{243}, a displacement of 2.5 mm which corresponds to an effective velocity of $V_y\approx 130$ km/s (the Alfv\'en velocity $V_A=70$ km/s).

These plasmoids were reproducible in their presence, but random in the time and position at which they formed.
Plasmoids did not form symmetrically around \y{0}, as would be expected if they were caused by initial density perturbations, but instead usually only one plasmoid was observed, in either the upper of lower half of the electron density map.
Plasmoids will be discussed further in \autoref{sec:discussion}.

\begin{figure}[!t]
	\Fig{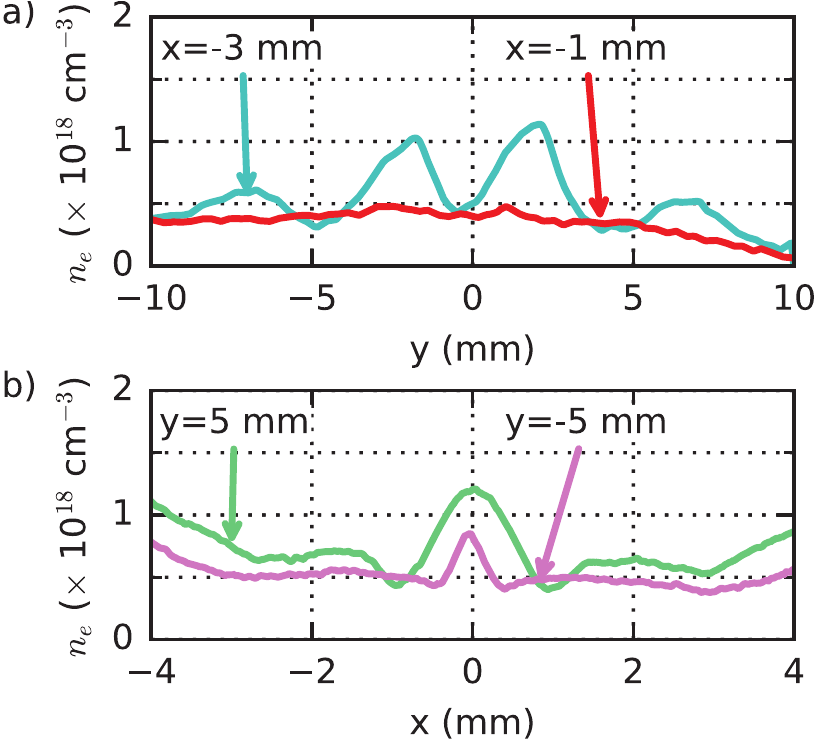}
	\centering
	\caption{Electron density line-outs from the maps in \hyperref[fig:electron_density_maps]{Fig. 3}. a) Line-outs of the inflow electron density profiles at \x{3} (blue) and \x{1} (red) from \hyperref[fig:electron_density_maps]{Fig. 3d}. b) Line-outs across the reconnection layer from \sref{fig:electron_density_maps}{e} --- note the density depletion at \y{5} (green) and \y{-5} (violet).}
\end{figure}

It is seen from \sref{fig:electron_density_maps}{} that plasma ablation is localised at the wires, and there is a strong azimuthal modulation close to the arrays.
As the flows are sub-Alfv\'enic, the outflows are smoothed as they propagate towards the layer, and we observe a good flow uniformity along the $y$-direction just outside the layer. 
This is seen in line-outs in \hyperref[fig:electron_density_maps]{Fig. 4a}, which correspond to the dashed blue and red lines at \x{-3} (blue) and \x{-1} (red) in \hyperref[fig:electron_density_maps]{Fig. 3d}.
The line-outs show that near the arrays (\x{-3}), the density in the incoming flows is strongly modulated, with $n_{e,max}/n_{e,min}\sim3$, where $n_{e,max}=$\xcmcubed{1.1}{18}.
Just outside the reconnection layer (\x{-1}), the modulations are negligible, with a roughly constant inflow density of \xcmcubed{0.9}{18}.

Representative line-outs across the reconnection layer are shown in \hyperref[fig:electron_density_maps]{Fig. 4b}, which correspond to the dashed lines at \y{5} (green) and \y{-5} (violet) in \hyperref[fig:electron_density_maps]{Fig. 3e}.
The density inside the plasmoid (\xcmcubed{1.2}{18}) is larger than in the rest of the reconnection layer (\xcmcubed{0.8}{18}).
It is also seen that there is a pronounced density depletion layer on either side of the reconnection layer, which is particularly evident around the plasmoid, where the density drops to \xcmcubed{0.4}{18}.

\begin{figure}[!h]
	\Fig{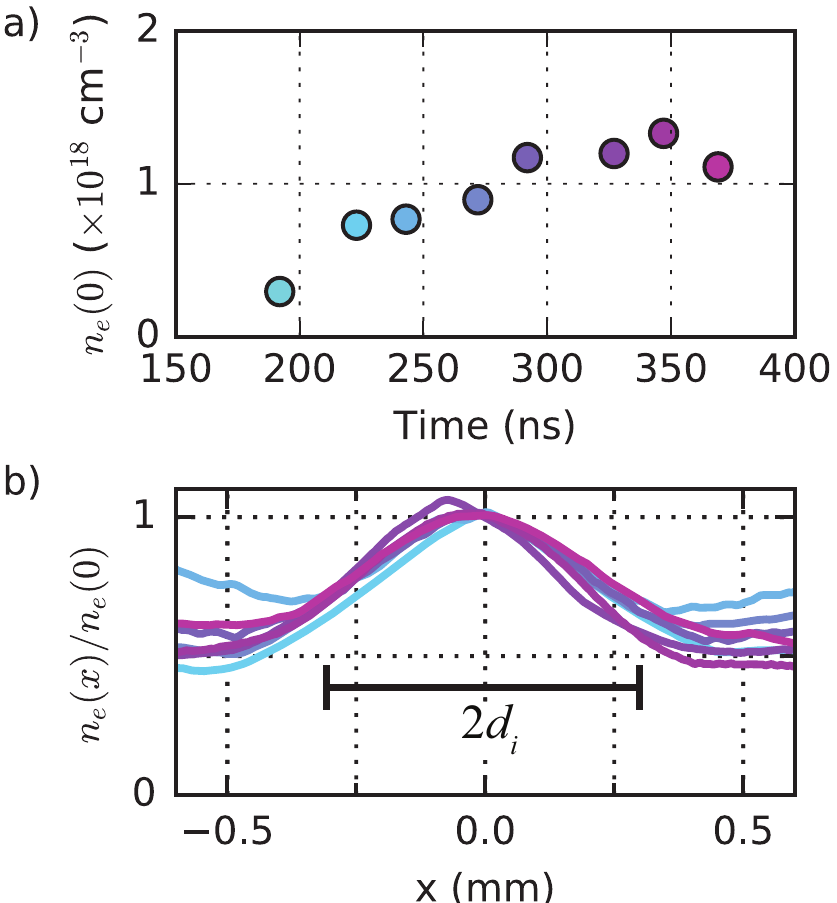}
	\centering
	\caption{a) Temporal evolution of the electron density in the centre of the reconnection layer (at $x$=0, $y$=0). b) Electron density profiles across the reconnection layer normalised to the density at the centre shown in a). Each profile is colour-coded and corresponds to the time with the same colour in a). The mean ion skin depth ($d_i=c/\omega_{pi}$) is shown on the same scale.}
\end{figure}

The evolution of the central electron density in the reconnection layer was determined using interferograms obtained at different times in different shots, and is shown in \hyperref[fig:electron_density_evolution]{Fig. 5a}.
The central electron density is taken as the average density in a square 0.2 mm wide, centered on the origin.
After an initial fast rise at the time of layer formation, the density plateaus and does not change appreciably.

Normalised electron density profiles across the reconnection layer are shown in \hyperref[fig:electron_density_evolution]{Fig. 5b}, which are line-outs of the electron density divided by the central electron density in \hyperref[fig:electron_density_evolution]{Fig. 5a}, and colour coded to show the time for each profile.
The profiles are remarkably similar, and show that the layer width does not change appreciably for over 150 ns.
The characteristic ion skin depth, $d_i=c/\omega_{pi}$ is shown on the same figure, and it is on the same order as the layer width, a necessary condition for the presence of two-fluid effects.

\begin{figure*}[t]
	\Fig{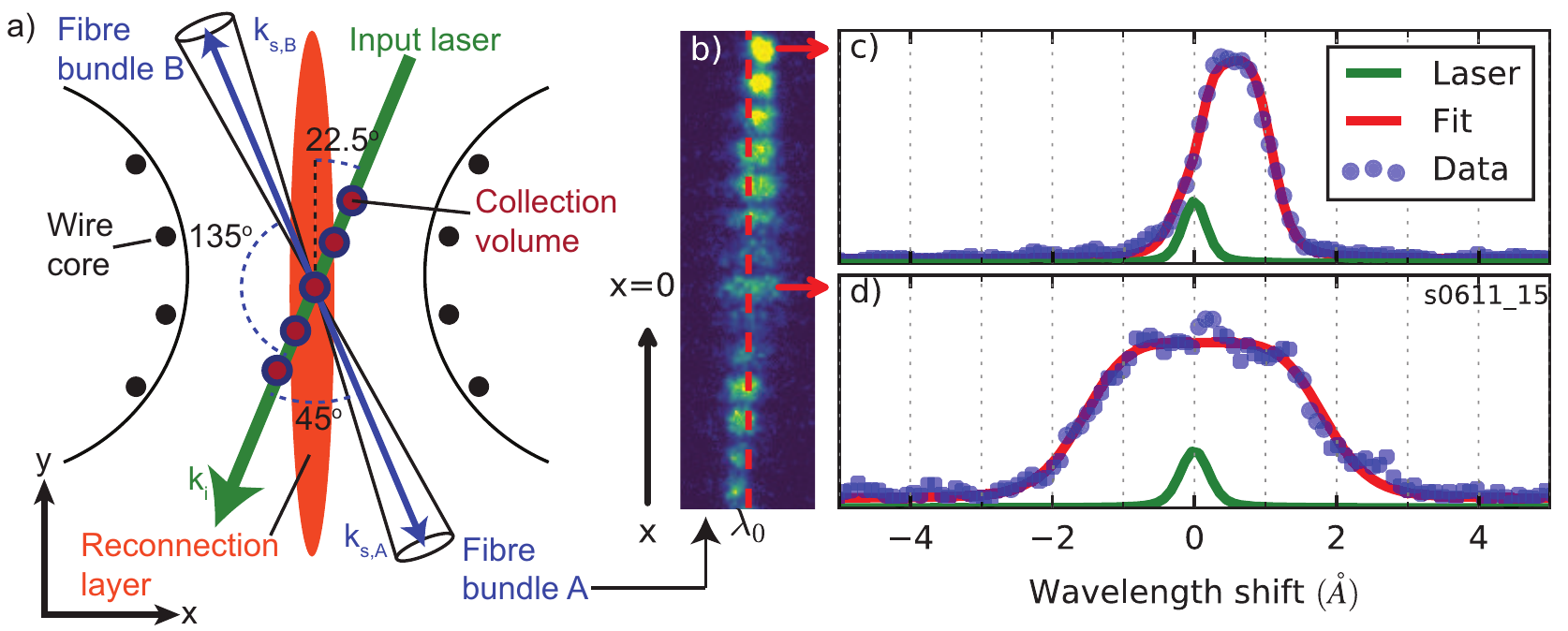}
	\centering
	\caption{Thomson scattering diagnostic set-up and sample data. a) Set-up of the Thomson scattering diagnostic, showing the probing laser beam passing through the reconnection layer and the two scattering vectors. b) Raw spectrogram from the fourteen fibre optics with scattering vector $\textbf{k}_{s,A}$, sensitive to Doppler shifts from velocity components in the $x$, or inflow, direction. The location of the fibre at $x=0$ mm is marked. c) Spectrum of scattered light (shown in blue) from outside the reconnection layer, fit with a theoretical form factor (red) and showing a significant Doppler shift and some Doppler broadening compared to the initial laser line (green). d) Spectrum of scattered light from the centre of the reconnection layer, showing a small Doppler shift and significant Doppler broadening.}
\end{figure*}

\subsection{Thomson Scattering}

The Thomson scattering geometry used in these experiments is shown in \hyperref[fig:thomson_scattering]{Fig. 6}.
The laser was focused using an f=2400 mm lens and propagted in the ($x,y$) plane through the reconnection layer at an angle of \deg{22.5} to the $y$-axis, crossing the layer close to the origin.
Light scattered from the plasma is Doppler broadened by thermal motion, and Doppler shifted by the bulk flow of the plasma, $\delta \omega=\textbf{k}\cdot\textbf{V}$, where $\textbf{k}=\textbf{k}_S-\textbf{k}_{in}$ is the difference between the vector defined by the observation angle $\textbf{k}_S$ and the laser beam $\textbf{k}_{in}$.
The two fibre optic bundles were at \deg{45} and \deg{135} to the probing laser beam (\hyperref[fig:thomson_scattering]{Fig. 6a}), and as such the spectra collected by these two bundles were sensitive to Doppler shifts from velocity components along the $x$ and $y$ axis respectively.

The scattered light was imaged onto the fibre optic bundles using two f=200 mm lenses (50 mm diameter) placed 440 mm from the reconnection layer, with the fibre bundles 367 mm away from the lenses.
This magnification ($M=0.83$) meant that the collection volumes were spaced by 450 $\mu$m along the laser beam, giving a total spatial range of 5.9 mm, with $\Delta x=2.3$ mm and $\Delta y=5.5$ mm.
The size of the collection volume was set by the intersection of the magnified image of the fibre optic (100/0.83 $\mu$m) and the diameter of the laser beam at the focal spot ($\sim200\,\mu$m), and so was approximately 200 $\mu$m.

A raw spectrogram taken at \t{232} after current start is shown in \hyperref[fig:thomson_scattering]{Fig. 6b}, showing data from fibre bundle A, which is at \deg{45} to the probing laser and hence sensitive to Doppler shifts in the inflow ($x$) direction.
The location of \x{0} is marked, and the spectra are Doppler shifted symmetrically about this point, implying that plasma flows symmetrically into the reconnection layer from both sides.
These spectra show more Doppler broadening close to the centre of the reconnection layer, indicating that the plasma has undergone significant heating.

Two sample spectra are shown in \hyperref[fig:thomson_scattering]{Fig. 6c} and \hyperref[fig:thomson_scattering]{Fig. 6d}. 
In \hyperref[fig:thomson_scattering]{Fig. 6c}, the scattered light comes from a volume outside the reconnection layer, at $x\approx1.1$ mm, and the spectrum is clearly Doppler broadened compared to the initial laser line, with a significant Doppler shift to longer wavelengths, indicating that the plasma flows in the $-x$ direction.
The scattered light in \hyperref[fig:thomson_scattering]{Fig. 6d} comes from the centre of the reconnection layer, close to \x{0}, and shows very little overall Doppler shift, but very significant Doppler broadening.
The shape of the spectrum here suggests that $\bar{Z}T_e\approx T_i$, with the two ion-acoustic peaks broadened to the extent that they form a flat top spectrum.\cite{Froula2011}

\begin{figure}[t]
	\Fig{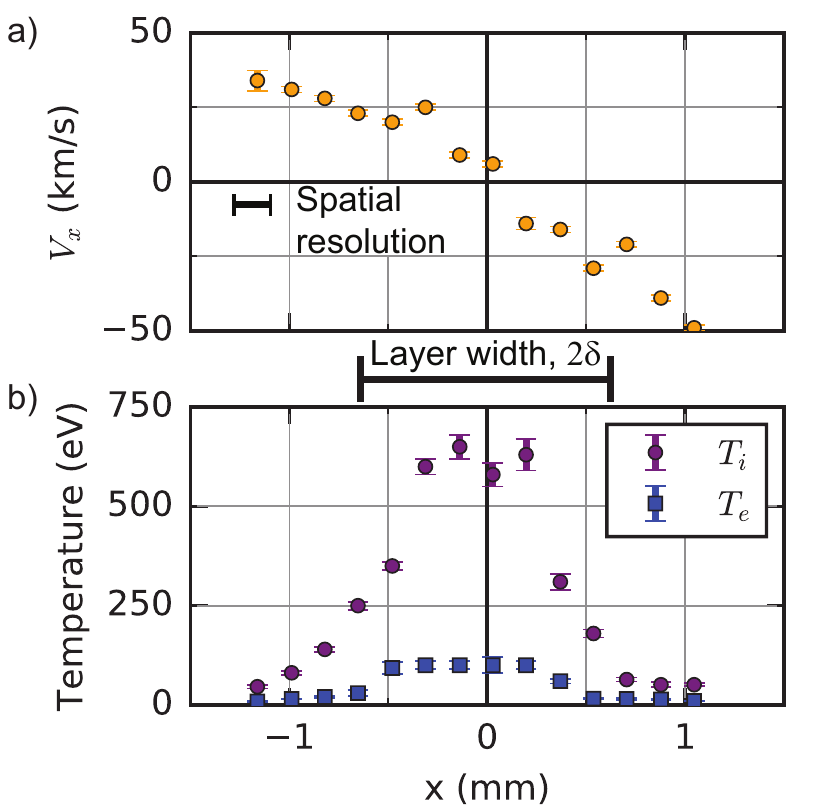}
	\centering
	\caption{Profiles of inflow velocity and electron ion temperatures obtained by fitting the data in \hyperref[fig:thomson_scattering]{Fig. 6b} at \t{232} after current start. The spatial resolution is set by the collection volume of the diagnostic, and the layer width is shown between the two profiles. a) Inflow velocity profile. b) Electron (blue squares) and ion (purple circles) temperatures.}
\end{figure}

Each of the fourteen spectra from distinct spatial locations shown in the spectrogram in \hyperref[fig:thomson_scattering]{Fig. 6b} were fit using theoretical form factors.\cite{Swadling2014a, Froula2011}
This produced the profiles of the inflow velocity, electron temperature and ion temperature shown in \hyperref[fig:thomson_profiles]{Fig. 6}.
The inflow velocity is initially around 50 km/s outside the reconnection layer (at $|x|=1$ mm \hyperref[fig:thomson_profiles]{Fig. 6a}), and the same velocity was measured further upstream in a separate shot at \x{3}, which indicates that there is very little change in velocity in the flows outside the reconnection layer.
The velocity drops to zero over a spatial scale of 1 mm, and follows a roughly linear decrease, except for the two data points near $|x|=0.5$ mm where the velocity sharply increases before dropping again.

The plasma was heated significantly over the same spatial scales as the flow decelerated (\hyperref[fig:thomson_profiles]{Fig. 6b}).
Initially the ion temperature is $T_i\leq50$ eV and $\zte\approx60$ eV.
This corresponds to $\bar{Z}=4$ and $T_e=15$ eV in nLTE.\cite{Chittenden2016}
As the flows approach the mid-plane, the ion temperature increases to over 600 eV, and the electrons are heated to $\zte{}\approx600$ eV, which corresponds to $\bar{Z}=6$ and $T_e=100$ eV, assuming the system has reached non-local thermodynamic equilibrium.

The outflow velocity was measured using two different scattering geometries. 
Close to the origin $V_y=50$ km/s, measured using the $B$ fibre bundle at \deg{135} to the probing laser, as shown in \hyperref[fig:thomson_profiles]{Fig. 6a}.
Further downstream at \y{6}, $V_y=130$ km/s was measured using a fibre optic bundle that observed the scattered light in the $+z$ (out-of-plane) direction, at \deg{90} to the probing laser which propagated along \x{0}.

The observed ion and electron heating was significant, and cannot be attributed simply to the thermalisation of the flow kinetic energy, as the kinetic energy $E_{kin}=Am_pV_{in}^2/2$ of a carbon ion ($A=12$) moving at 50 km/s is 100 eV, far less than the $600 eV$ measured in the reconnection layer.
Instead, we consider the role of the magnetic energy in heating the plasma, and hence look for evidence of flux annihilation, which is one of the principle signatures of magnetic reconnection.

\subsection{Magnetic field measurements}

\begin{figure}[t]
	\Fig{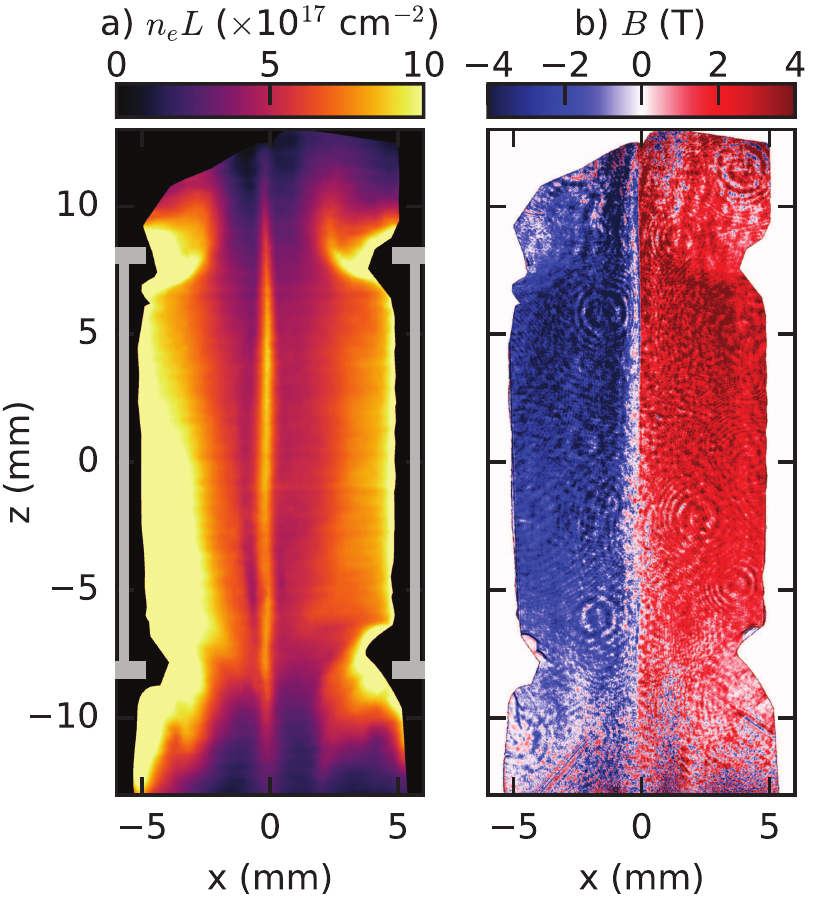}
	\centering
	\caption{Results from the Faraday rotation imaging diagnostic, taken at \t{251} after current start. a) Side on areal electron density map (location of wires and array hardware shown in grey). b) Reconnecting magnetic field ($B_y$) map, same time and spatial scale as in a).}
\end{figure}

Magnetic field distributions were measured using Faraday rotation imaging, with probing performed in the $y$-direction, offering a side-on view of the reconnection layer, as in \sref{fig:3D_double_exploders}{c}.
Data from the Faraday rotation imaging diagnostic is shown in \sref{fig:faraday}, which views the reconnection layer side-on (as in \hyperref[fig:3D_double_exploders]{Fig. 2c}), so that the reconnecting magnetic field points into and out of the page.
The areal electron density map from the in-line interferometry diagnostic is shown in \sref{fig:faraday}{a}, with the location of the two arrays on the left and the right sides overlaid in grey.
The reconnection layer is clearly visible at the centre of the image, with the characteristic density depletion region just outside the layer that was shown in \hyperref[fig:electron_density_maps]{Fig. 3d}.
Aside from the slight left/right asymmetry near the bottom of the array, the layer has a clear up/down symmetry which justifies using a quasi-2D assumption in our analysis.

This areal electron density map is combined with the Faraday rotation map, or polarogram (Fig. 3a, Ref. \onlinecite{Hare2017}), to produce the reconnecting magnetic field map shown in \sref{fig:faraday}{b}.
The magnetic field map is colour coded --- blue indicates magnetic fields pointing out of the page, red shows fields pointing into the page, whilst white indicates no magnetic field.

\begin{figure}[!h]
	\Fig{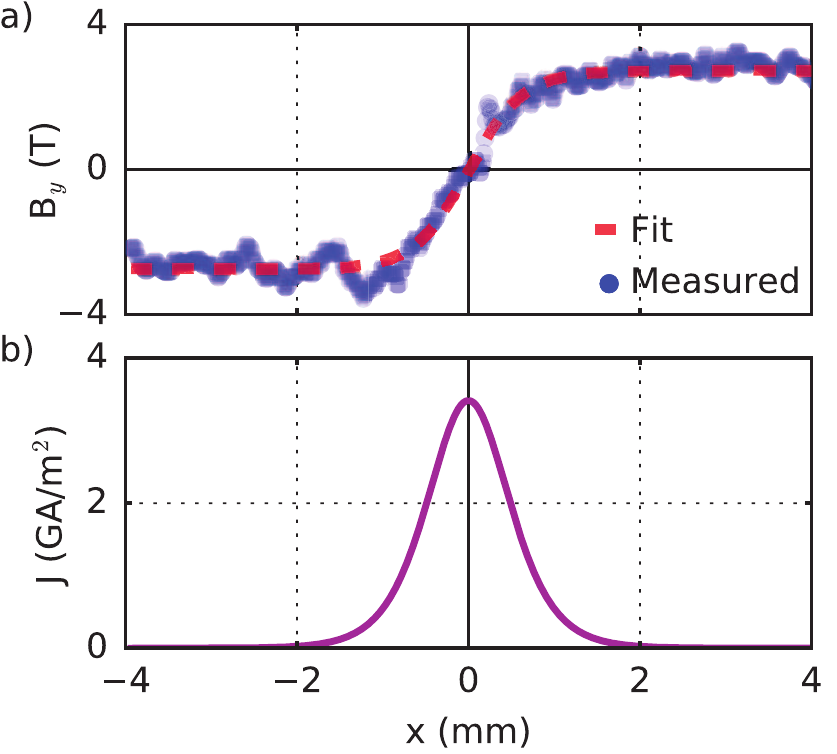}
	\centering
	\caption{a) Line-out of the magnetic field (blue) from \sref{fig:faraday}{b}, taken along \z{0\pm2}, and the Harris sheet fit (red). b) Electric current density inferred from the Harris sheet in a).}
\end{figure}

The reconnecting magnetic field map shows excellent up/down symmetry as well as left/right inversion symmetry.
A line-out of the reconnecting magnetic field is shown in \sref{fig:harris}{a}.
This profile is well approximated by the Harris sheet profile\(B_y(x)=B_0\tanh{\left(\delta/L\right)}\)\cite{Harris1962}, with the reconnecting field strength \(B_0=2.9\) T and the layer half-width \(\delta=0.6\) mm.
This layer half-width is in agreement with the results from the electron density maps (\hyperref[fig:electron_density_maps]{Fig. 3d}).

\begin{figure*}[t]
	\Fig{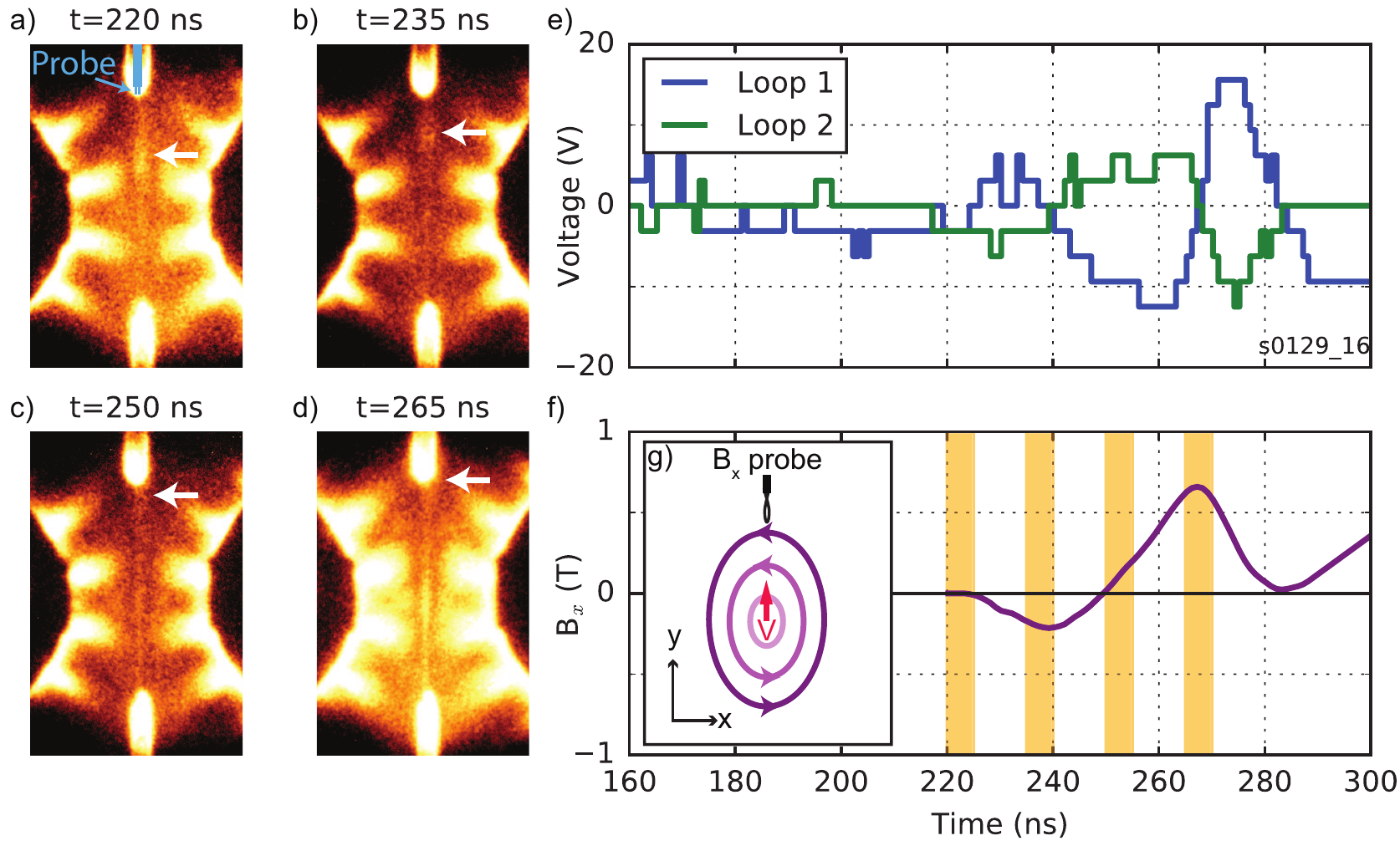}
	\centering
	\caption{a)-d) Images obtained using optical fast-framing camera (5 ns exposure, $\lambda>600$ nm). Two sets of probes are visible [one set drawn to scale in purple at the top of the fast-framing image in a)], and these probes are sensitive to the reconnected ($B_y$) component of the magnetic field. In each frame a plasmoid, marked with a white arrow, can be seen colliding with the top probe which consists of two oppositely wound loops, and the voltage signals from these two loops are shown in e). Prior to the arrival of the plasmoid the signal is very small, indicating a small reconnected magnetic field. f) The signals are combined to remove the electro-static component, and integrated to give an oscillating magnetic field consistent with the O-point magnetic field line structure of the plasmoid, shown in g). The time of the four frames a)-d) are shown as orange bars in f). \href{https://goo.gl/05tzBU}{(Multimedia view)}.}
\end{figure*}

Magnetic probes were used to measure the reconnected magnetic field and the magnetic field structure of the plasmoids.
Two pairs of probes were placed in the outflows of the reconnection layer, with the probe loops oriented so that they were sensitive to the $B_y$ component of the magnetic field.
The probes were within the field of view of the end-on probing of the laser interferometry and the optical fast-framing camera, as seen in \sref{fig:bdots}{a--d}.
The signal from the top pair of probes is shown in \sref{fig:bdots}{e}.
There was very little signal before around \t{220}, which indicates that the reconnected magnetic field strength was negligible.

Plasmoids were identified within the reconnection layer using the optical fast-framing camera (see Supplemental Material \href{https://goo.gl/OjqA4M}{here} for an animation from a shot without magnetic probes, or \sref{fig:bdots}{a--d} for images from a shot with magnetic probes).
This camera provides twelve frames of data per shot, and this broad temporal range allows us to capture plasmoid formation, which varies in time and space from shot to shot. 
Plasmoids were observed in each of the eight shots in which the optical fast-framing camera was used.

In \sref{fig:bdots}{a--d}, one plasmoid has been marked with a white arrow --- this plasmoid travelled outwards along of the reconnection layer and interacted with the top probe, and the signal of the probe can therefore be correlated with the plasmoid's location.
These probes use two oppositely wound loops without any built-in integration, which allows the fidelity of the signal to be determined by examination of the raw voltage signals in \sref{fig:bdots}{e}.
The signals from the two loops are similar, with opposite signs, which suggests that the electrostatic component is negligible --- as such the integration represents a true reconstruction of the magnetic field rather than the integration of a spurious signal.

The integrated magnetic field is shown in \sref{fig:bdots}{f}.
The integration begins when the signals in \sref{fig:bdots}{e} exceeded the noise threshold of the oscilloscope, at around \t{220}.
The times of the four frames (5 ns exposure, 15 ns between frames) are shown as orange bars in \sref{fig:bdots}{f}.
In the first frame (\sref{fig:bdots}{a}, \t{220}), the plasmoid had not yet reached the probe and the magnetic field was small.
As the plasmoid passed through the probe (\sref{fig:bdots}{b}, \t{235}), the magnetic field increased and then fell as the centre of the plasmoid passes through the probe (\sref{fig:bdots}{c}, \t{250}).
The magnetic field then increased again with the opposite sign as the plasmoid left the region where the probe measured the magnetic field (\sref{fig:bdots}{d}, \t{265}), before falling back to zero.
This profile is consistent with the O-point structure predicted for plasmoids.

It is important to note that the magnetic probes occupy only 0.5 mm in the $z$ direction, which is small compared to the overall height of the reconnection layer of 16 mm.
If the plasmoid is long compared to the probe then the probe should not disturb the overall structure of the magnetic field distribution.
The plasmoids are seen in the end-on interferograms, which suggests that the vertical extent of the plasmoid is indeed long, and comparable with the height of the reconnection layer.

\section{Discussion}\label{sec:discussion}

The detailed measurements presented in the previous section allow us to make quantitative comparisons of our results with theories of magnetic reconnection.
We begin by presenting a comparison of the observed inflow and outflow velocities with the predictions of a generalised Sweet-Parker model which includes compressibility effects, pressure balance and ionisation inside the layer\cite{Ji1999}.
We consider the power balance between the inflows and outflows, which shows a good overall agreement and emphasises the important role of the magnetic energy in heating the ions and electrons.
However, time-scales for classical heating mechanisms are found to be too long to explain the high electron and ion temperatures, which suggests that other, more efficient mechanisms must be at work.
We consider the reconnecting electric field and its components under a steady-state, resistive MHD assumption, and then discuss what mechanisms might account for the imbalance between the convective and resistive terms.
The formation of plasmoids within the reconnection layer is considered as a mechanism for anomalous heating and the electric field imbalance, and we make comparisons of our observations to the predictions plasmoid instability theory.

\subsection{Comparison with the generalised Sweet-Parker model}
One of the most enduring models of magnetic reconnection is due to Sweet\cite{Sweet1958} and Parker\cite{Parker1957}, which assumes a steady state system, and uses the conservation of mass and a pressure balance to predict a reconnection rate of \(V_{in}/V_A=S^{-1/2}\), where $V_A$ is the Alfv\'en velocity and $S=\mu_0 V_A L/\eta$ is the dimensionless Lundquist number (and $\eta/\mu_0$ is the magnetic diffusivity.)
In this experiment $S\approx120$ (using the parameters in \autoref{tab:params}), and so $V_{in}\approx V_A/10\approx 7$ km/s, far slower than the observed inflow velocity of $50$ km/s.
Additionally, the outflow speed in the Sweet-Parker model is set at $V_A$ due to pressure considerations, but we observe outflows at $V_{out}=130$ km/s, far larger than $V_A=70$ km/s.
These discrepancies suggest that more complicated physics needs to be added to the Sweet-Parker model in order to account for our observations.

The Sweet-Parker model was modified by Ji et. al. \cite{Ji1999} to include the effect of compressibility and the pressure downstream of the outflows.
In this generalised Sweet-Parker model, the outflow velocity can be enhanced or reduced by the pressure difference between the reconnection layer and the environment into which the outflows expand, giving:
\begin{equation}
V^2_y=V_A^2-\frac{2(p_{down}-p_{up})}{m_in_i}
\end{equation}
In contrast to MRX, where the downstream pressure is significant and slows the outflows\cite{Ji1999}, in our experiment the outflows expand into a vacuum with $p_{down}=0$, and $p_{up}/\rho=k_B(\zte+T_i)/m_i=C_{i,A}^2$, where $C_{i,A}$ is the ion-acoustic velocity.
This means that:
\begin{equation}
V_y=\sqrt{V_A^2+2C_{i, A}^2}= 140 \pm 4\,\textrm{km/s},
\end{equation}
which is in good agreement with the outflow speed inferred from the plasmoid motion and measured using Thomson scattering.

In our experiment, we need to further modify this generalised Sweet-Parker model to include the effects of ionisation inside the reconnection layer, so that the conservation of mass is explicitly applied to the ion density.
By modifying eqn 5. of Ji et. al. \cite{Ji1999} we obtain for the inflow velocity:
\begin{equation}
\label{eqn:compressibility}
V_x=\frac{\delta}{L}\left( V_y\frac{n_2}{n_1}+\frac{L}{n_1}\frac{\partial n_2}{\partial t}\right)=31\pm 4\, \textrm{km/s},
\end{equation}
where \(n_1\) is the ion density at the edge of the layer (\(|x|=0.6\) mm) and \(n_2\) is the ion density at the centre of the layer (\(x=0\) mm). 
The ion densities are calculated using the electron density map in \hyperref[fig:electron_density_maps]{Figs. 3a} and the values for $\bar{Z}$ determined from Thomson scattering using the nLTE model.
The increase in density due to compressibility effects (\(\partial{}n_2/\partial{}t\)) is determined using \hyperref[fig:electron_density_maps]{Figs. 3a and 3b} which were taken 20 ns apart in the same experiment.
The time between the two frames is significantly shorter than the flow transit time out of the reconnection layer ($\tau_{exp}=L/V_{out}\approx 50$ ns).

The generalised Sweet-Parker model gives predictions which are consistent with our results, demonstrating that compressibility, ionisation and downstream pressure are important elements in determining the reconnection rate in these experiments.

\subsection{Power balance and anomalous heating} 

\begin{figure}[h]
	\Fig{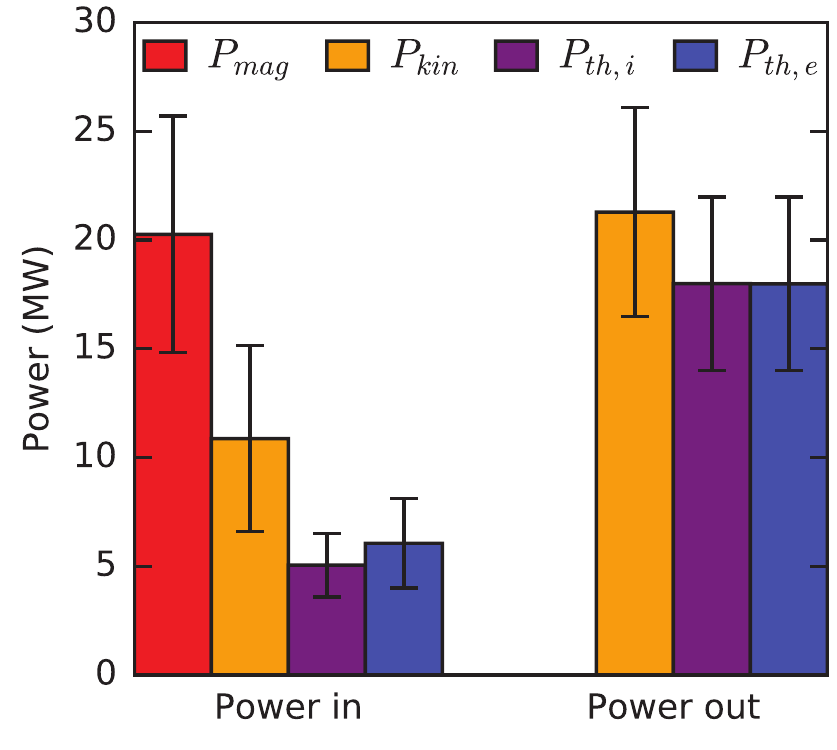}
	\centering
	\caption{Power balance for power flowing into and out of the reconnection layer. The power in (out) is calculated by multiplying the energy density for each component by $V_{in} L h$ ($V_{out} \delta h$). In the out flow, the magnetic field energy is assumed to be negligible, which is supported by the data in \sref{fig:bdots}{f}.}
\end{figure}

The power flowing into and out of the reconnection layer can be calculated using the detailed measurements of the plasma parameters presented in the previous section.
We consider four components, the magnetic energy density $E_{mag}=B^2/2\mu_0$, the kinetic energy density $E_{kin}=m_{i}n_{i}V_{in}^2/2$, the ion thermal energy density, $E_{th, i}=3 n_{i}k_B T_{i}/2$ and the electron thermal energy density, $E_{th,e}=3 n_{e}k_B T_{e}/2$ .
These energy densities are calculated for the inflow and outflow conditions shown in \hyperref[tab:params]{Table I}, and multiplied by the volume of plasma that enters or exits the reconnection layer per unit time: $LhV_{in}$ for the inflows and $\delta h V_{out}$ for the outflows, where $h=16$ mm is the height of the experiment.

The components of the inflow and outflow power are shown in \sref{fig:power_balance}, with error bars derived from the experimental uncertainties of the measurements.
The overall power balance between the inflows and outflows is good, with the sum over the energy components in agreement within experimental uncertainty.
For the inflows the magnetic energy clearly dominates over the other energy components, and as such plays a significant role in heating the plasma.
We neglect the reconnected magnetic field energy density in the outflows --- the standard Sweet-Parker model predicts $E_{mag, out}=E_{mag, in}/S\approx0.01E_{mag, in}$ (far smaller than the experimental uncertainty) and the magnetic probes \sref{fig:bdots}{f} also showed a negligible reconnected field, both of which justify this assumption.
We note that in a collisionless system, recent experimental work has shown that the outflow electromagnetic energy is significant\cite{Yamada2015a, Yamada2016}, which differs from the result we obtain in collisional reconnection.
It is seen in \sref{fig:power_balance}{} that the magnetic energy is converted almost equally into the remaining energy components, with the outflow kinetic energy of the flows slightly greater than the electron and ion thermal components, which are equal.

Although the thermal energy components of the electrons and ions are roughly equal, the temperatures are significantly different, with $T_i\approx6T_e$.
The ions and electrons can maintain different temperatures due to infrequent collisions, as the electron-ion equilibration time-scale is far longer than the experimental time-scale,  $\tau_{E, ei}\approx250\,\textrm{ns} \gg\tau_{exp}=L/V_{out}\approx 50\,\textrm{ns}$.\cite{Ryutov2015a}

As the time-scale for temperature equilibration is large, we can calculate the classical heating rates separately for the ions and electrons, as follows.
For the ions, we consider viscous damping of the highly sheared velocity profile of the outflows, and following the treatment by Hsu et al.\cite{Hsu2001a} we solve:
\begin{equation}
\label{eqn:viscous_heating}
\frac{3}{2}n_ik_B\frac{\partial T_i}{\partial t}=0.96 n_i k_B T_i \tau_i \left(\frac{\partial V_y}{\partial x}\right)^2,
\end{equation}
where we have used Braginskii's expression for the ion viscosity,\cite{Braginskii1965} with $\tau_i\propto T_i^{3/2}$.
We assume that $V_y$ drops to zero outside the layer, giving a velocity shear of $\partial V_y/\partial x\approx V_y/\delta$, and use the parameters in \hyperref[tab:params]{Table I}.
Solving eqn. \ref{eqn:viscous_heating} gives $\tau_{visc}=800$ ns, the time-scale for viscous heating of the ions from their initial temperature in the flow of 50 eV to the 600 eV observed in the layer.
This is much longer than the experimental time-scale, $\tau_{exp}=50$ ns, and we conclude that the classical viscous heating rate is too slow to explain the high ion temperatures we observe.

A similar calculation can be performed for the electrons.
We use an non-local thermodynamic equilibrium model\cite{Chittenden2016} to calculate the radiative cooling time scale for the electrons as $\tau_{rad}=600$ ns, which means that radiative cooling is negligible on the experimental time-scale and we can consider only Ohmic heating, solving:
\begin{equation}
\label{eqn:ohmic_heating}
\frac{3}{2}n_ek_B\frac{\partial T_e}{\partial t}=\eta_{Sp}J^2,
\end{equation}
where $\eta_{Sp}$ is the classical Spitzer-Braginskii resistivity, $\eta_{Sp}\propto T_e^{-3/2}$.
The solution of eqn. \ref{eqn:ohmic_heating} predicts it would take $\tau_{res}=350$ ns for the electrons to heat from 15 eV to 100 eV, which shows that classical resistive heating is not an efficient mechanism on our experimental time-scale.

Both the electron and ion heating is therefore anomalous, in that it cannot be explained by classical mechanisms.
Similar anomalous heating has been observed in many other experiments, and a range of mechanisms have been proposed to account for the discrepancy, including anomalous resistivity and anomalous viscosity\cite{Hsu2001a} due to kinetic turbulence, or in-plane electric fields caused by two-fluid effects.\cite{Yoo2014}.
These heating mechanisms have additional signatures, and we see further evidence for the need to include physics beyond resistive MHD when we consider the reconnecting electric field in the next section.

\subsection{Electric field balance across the current sheet}

The reconnecting electric field points out of the reconnection plane (in the $+z$ direction), and sets the rate of reconnection through flux annihilation.
In a steady state system, the reconnecting electric field is constant across the reconnection layer, and is supported by different components of the generalised Ohm's law in different regions.
For resistive magnetohydrodynamics (MHD), Ohm's law across the reconnection layer is:
\begin{equation}
\label{eqn:rmhd_ohms_law}
E_z+V_x\times B_y=\eta_{Sp}J_z,
\end{equation}
and so outside the reconnection layer, where $J_z=0$, the electric field is supported by the convective term $V_x\times B_y$.
Inside the layer, where the magnetic field and inflow velocity are small, the electric field should be supported by the resistive term $\eta_{Sp}J_z$, where $\eta_{Sp}$ is the classical Spitzer-Braginskii resistivity.

\begin{figure}[h]
	\Fig{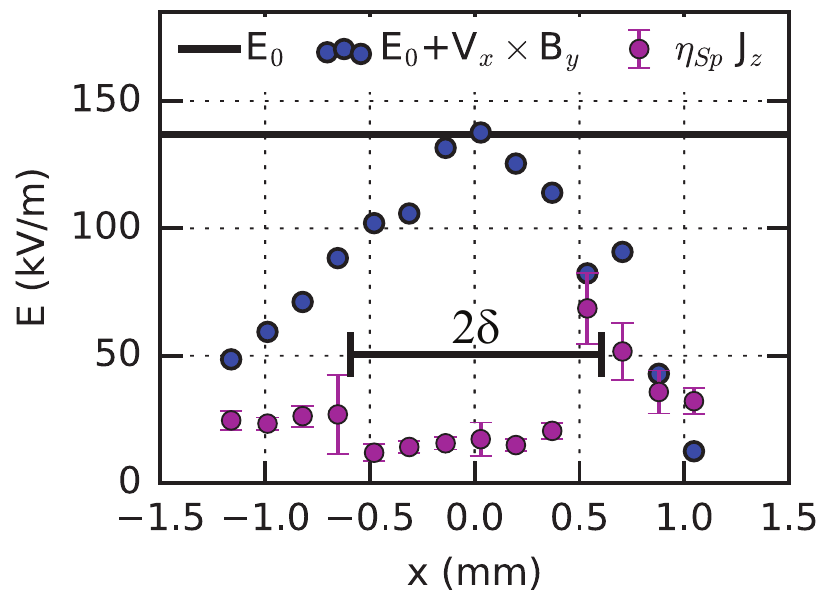}
	\centering
	\caption{Components of the reconnecting electric field ($E_z$) across the reconnection layer, showing the constant electric field $E_0$ predicted by the steady state assumption, the sum of $E_0$ and the convective electric field $V\times B$, and the resistive component of the electric field $\eta_{Sp}J$, calculated using the classical resistivity.}
\end{figure}

Using our experimental results, we can calculate the these components and see whether this simple Ohm's law is balanced in our reconnection layer (\hyperref[fig:electric_field]{Fig. 10}).
We calculate the current profile using the Harris sheet profile, $J_z=(B_0/\mu_0\delta)\sech^2\left(x/\delta\right)$ (\sref{fig:harris}{b}), where $B_0$ and $\delta$ are determined from Faraday rotation imaging.
In \sref{fig:electric_field}, the constant electric field predicted in steady state ($E_0=V_{x,0}\times B_{y,0}=140$ kV/m, using $V_{x,0}=50$ km/s and $B_{y,0}=2.8$ T measured outside the layer) is shown as a black line.
The left hand side of eqn. \ref{eqn:rmhd_ohms_law} ($E_0+V_x\times B_y$) is shown with blue dots, and equals $E_0$ at \x{0} where $V_x$ and $B_y$ both go to zero.
The resistive component (the right hand side of eqn. \ref{eqn:rmhd_ohms_law}, $\eta_{Sp}J_z$) is shown in purple, and should be equal to $E_0+V_x\times B_y$ everywhere if steady-state, resistive MHD holds for this system.
Instead, it is seen that the resistive term is too small almost everywhere in the layer, and especially at the mid-plane where it is a factor of $\sim10$ smaller than necessary to support the reconnecting electric field.

There are several possible reasons why the electric field components fail to balance in these experiments:
Firstly, there could be additional terms in Ohm's law, caused by, for example, two-fluid effects, which create in-plane currents that in turn produce a characteristic quadrupolar magnetic field structure, discussed in Ref. \onlinecite{Zweibel2016} and references therein.
These in-plane currents contribute to the reconnecting electric field through the Hall term, $\textbf{J}\times\textbf{B}/en_e$, which is known to play a dominant role in collisionless plasmas.\cite{Fox2011}
The layer width $\delta$ in our experiments is comparable to the inertial length, $d_i=c/\omega_{pi}$, a necessary condition for the presence of two-fluid effects, and we observe a density depletion region (see \sref{fig:electron_density_lineouts}{b}) which has been associated with two-fluid effects in simulations\cite{Shay2001} and satellite observations.\cite{Mozer2002}
Conversely, the short mean-free-path in our experiments ($L/\lambda_{ii}\sim200$) suggests that collisional effects should dominate, and preliminary work has shown no evidence for the presence of a quadrupolar magnetic field structure.
Clearly further investigation is necessary to quantify what role two-fluid effects may have in our magnetic reconnection experiments.

Secondly, the resistive term could be larger than the Spitzer-Braginskii value if it is enhanced by kinetic turbulence, which could increase the collisionality through wave-particle interactions.
The resistivity would need to be enhanced by a factor of ten in order to support the reconnecting electric field using only the resistive term at the centre of the layer.
This factor of ten enhancement would also reduce the  Ohmic heating time-scale to less than the experimental time-scale, which suggests that anomalous resistivity could provide the enhanced heating mechanism necessary to explain the high electron temperatures we observe.
Using the measured plasma parameters in \autoref{sec:results}, we find that \(T_i\approx\bar{Z}T_e\) and \(C_{i,A}\approx{}u_{ed}\) (\(u_{ed}\) is the electron drift velocity, \(j=en_{e}u_{ed}\)) which are common criteria for the development of instabilities such as the ion-acoustic or lower-hybrid drift.
In other experiments evidence for kinetic turbulence has come from the observation of high frequency waves\cite{Carter2001,Ji2004}, measured using in situ magnetic probes.
In our conditions, one could attempt to detect such waves using Thomson scattering\cite{Gray1980,Glenzer2002} and we have begun preliminary work to look for the presence of turbulence.

A third possibility is the breaking of the steady-state assumption which implies a constant electric field.
The magnetic field does change gradually during the experiment due to the rising current pulse, but this change is not fast enough to explain the discrepancy in the electric field balance seen in \sref{fig:electric_field}.
However, the plasmoids are one source of rapidly changing magnetic fields, as shown in \sref{fig:bdots}{f}.
The fast motion and O-point structure of the plasmoids should break the steady-state assumption, leading to a reconnecting electric field which varies over the reconnection layer.
The role of the plasmoids in balancing the electric field will be the subject of future work.
Below we present more discussion of the plasmoids observed in these experiments.

\subsection{Plasmoids}\label{ssec:plasmoids}

In our experiments, plasmoids were observed independently using laser interferometry and an optical fast-framing camera (see Supplemental Material \href{https://goo.gl/OjqA4M}{here} for an animation comprising of 12 frames).
The plasmoids formed within the reconnection layer at random locations and moved outwards along the layer at super-Alfv\'enic velocities.

We emphasise that these plasmoids are not formed by the initial density perturbations of the inflows:
if they were, then by symmetry we would expect an equal number of plasmoids at symmetric locations on both sides of \y{0}, but we usually only observe a single plasmoid.
Additionally, the inflows were not modulated by the time they reached the mid-plane, as shown in \hyperref[fig:electron_density_maps]{Fig. 3c} --- this could be due to the sub-Alfv\'enic nature of the inflows, which allows the density variations to be smoothed out.
Finally, the plasmoids clearly have an O-point magnetic structure, as demonstrated by the in-situ measurements presented in \sref{fig:bdots}, confirming that these plasmoids are more than hydrodynamic artefacts of the initial conditions.

Plasmoids have also been observed in other recent laboratory experiments on TREX\cite{Olson2016} and MRX\cite{Jara-almonte2016}, but there is no theory which predicts the properties of plasmoids in the regimes in which these experiments occur.
Using the measured plasma parameters, we find that the plasmoids we observe are consistent with the semi-collisional regime of the plasmoid instability\cite{Loureiro2015}, defined in the asymptotic limit as $(L/d_i)^{8/5}\ll S \ll (L/d_i)^2$, where in our plasma $L/d_i=18$ and $S=120$.
In the collisional regime of the plasmoid instability, which most theoretical and numerical work has focused on, there is a critical Lundquist number $S_c=10^4$, below which plasmoids cannot form.
This critical Lundquist number is absent in the semi-collisional regime\cite{Loureiro2015}, and so plasmoids can form even at modest Lundquist numbers specified by the condition above, which is more easily achieved in simulations and laboratory experiments.

We can compare our observations with the predictions of the linear theory of semi-collisional plasmoids, in which the growth time of the  instability is estimated as \((L/d_i)^{6/13}S^{-7/13}L/V_A\sim30\) ns and the number of plasmoids predicted is \((d_i/L)^{1/13}S^{11/26}/2\pi\sim3\).\cite{Baalrud2011}
The prediction for the linear growth rate suggests that there is sufficient time in our experiments for the plasmoids to enter the non-linear regime, which is the regime in which we can observe them.
With regards to the number of plasmoids, since linear theory predicts relatively few of them, plasmoid coalescence is not supposed to be play an important role, and so the relative agreement between the linear prediction and our observations is encouraging. 

For plasmoids in the collisional regime, it is known that the production of multiple X-points and highly sheared flows provide an enhancement to the resistive and viscous heating rates.\cite{Loureiro2012}
So far there have been no numerical or theoretical studies to determine whether this effect carries over to the semi-collisional regime in which the plasmoids we observe exist.
If enhanced heating is found to occur in this regime, it could offer another explanation for the enhanced electron and ion temperatures observed.

\section{Conclusions}\label{sec:conclusions}

We present the results from a recently developed pulsed-power driven reconnection experiment, which produces a symmetric reconnection layer from super-sonic, sub-Alfv\'enic inflows with $\beta_{th}\sim\beta_{dyn}\sim1$.
This layer is well-diagnosed by a suite of non-perturbative laser based diagnostics which provide temporally and spatially resolved measurements of the electron density, magnetic field, flow velocity and electron and ion temperatures.
We observe the formation of a long lasting layer with a normalised density profile which remains constant for over 150 ns, surrounded by a density depletion region.
A dramatic increase in the electron and ion temperatures was observed using Thomson scattering, and this increase was larger than can be explained by the thermalisation of the kinetic energy in the inflows.
Faraday rotation imaging showed a symmetric, quasi-2D reconnection layer which is well approximated by the Harris sheet, with a width of $\delta=0.6$ mm, and hence an aspect ratio of $L/\delta\sim10$.

Using laser interferometry and a fast framing camera, we observe the repeated formation of plasmoids which are ejected along the the layer at super-Alfv\'enic velocities.
In-situ measurements using magnetic probes show negligible reconnected magnetic field, consistent with predictions for collisional reconnection, and these measurements also demonstrate that the plasmoids have an O-point magnetic field structure.

We showed good agreement between the reconnection rate predicted by the generalised Sweet-Parker model and our observations, and this emphasises the importance of compressibility and the open boundary conditions in determining the high inflow and outflow velocities in this experiment.
There was good overall power balance in the reconnection layer, in which the magnetic energy flowing into the layer was converted to thermal and kinetic energy in the outflows.
Classical viscous and resistive mechanisms are far too slow to efficiently heat the ions and electrons, implying that other mechanisms such as two-fluid effects, kinetic turbulence or heating by plasmoids must be considered.

Using a steady state assumption, the electric field balance across the reconnection layer was calculated, and the classical resistive term was found to be too small to support the reconnecting electric field.
This in turn suggested the presence of two-fluid effects, enhanced resistivity through kinetic turbulence, or rapid magnetic field fluctuations caused by plasmoids, all of which will be the subject of further investigations with this experimental platform.

\section*{Acknowledgements}

This work was supported in part by the Engineering and Physical Sciences Research Council (EPSRC) Grant No. EP/N013379/1, and by the U.S. Department of Energy (DOE) Awards No. DE-F03-02NA00057 and No. DE-SC-0001063. AC was supported by LABEX Plas@Par with French state funds managed by the ANR within the Investissements d'Avenir programme under reference ANR-11-IDEX-0004-02. NFL was supported by the NSF-DOE partnership in basic plasma science and engineering, award no. DE-SC0016215

\bibliography{library}

\end{document}